\documentclass[runningheads]{svmult}

\def\hwplusplus{\textsf{Herwig++}}
\def\herwig{\texttt{HERWIG}}
\def\pythia{\texttt{Pythia}}
\def\p7{\textsf{Pythia7}}
\def\amegic{\textsf{AMEGIC++}}
\def\apacic{\textsf{APACIC++}}
\def\madevent{\texttt{MadEvent}}
\def\alpgen{\texttt{AlpGen}}
\def\comphep{\texttt{CompHEP}}
\def\geant4{\textsf{geant4}}
\def\EvtGen{\textsf{EvtGen}}
\def\mcnlo{\texttt{MC@NLO}}
\def\lqcd{\Lambda_{\mbox{\tiny QCD}}}
\def\as{\alpha_{\mbox{\tiny S}}}

\usepackage{epsfig,bm,picins}
\usepackage{graphicx}  
\usepackage{physmubb}  

\begin{document}
\title*{Event Generators --- New Developments
   \thanks{Invited talk
     at the \emph{14th Topical Conference on Hadron Collider
       Physics} (HCP2002), Karlsruhe, Germany, 30 Sep--4 Oct 2002, to
     appear in the proceedings. }
   \unitlength1pt
   \begin{picture}(0,0)
     \put(-62,30){\footnotesize\rm Cavendish-HEP 02/19}
   \end{picture}
}
\toctitle{Event Generators --- New Developments}
\titlerunning{Event Generators --- New Developments}
%
\author{Stefan Gieseke \\
\textit{Cavendish Laboratory, University of Cambridge \\
  Madingley Road, Cambridge CB3 0HE, U.K. }}
\authorrunning{Stefan Gieseke}

\maketitle              

\section{Introduction: event generators}

As in the past, Monte Carlo event generators will play a vital r\^ole
for the physics analysis of events at present and future hadron
colliders like the Tevatron, HERAII and, of course, the LHC.  The
particular advantages of the simulation of single events with a
multi-purpose event generator like \herwig\ \cite{Herwig6,Herwig65} or
\pythia\ \cite{Pythia} are all related to the fact that a fully
exclusive final state is obtained.  This allows to run the output
through the same analysis tools as the data that was measured, thereby
allowing to apply e.g.\ the same physical cuts as those applied to the
data \emph{after} the generation of events.  This means that the
events generated by the Monte Carlo program are treated in exactly the
same way as the data from the measurement.

In this overview we revisit the basic facts and features of
multi-purpose event-generators.  In the following we consider some
recent approaches to a significant improvement of the quality of event
generators with the help of matrix elements: methods to match the
results of matrix element calculations with those of the parton shower
approach are discussed in Sec.~\ref{sec:meps}.  Finally, in
Sec.~\ref{sec:hw++} we describe the ongoing development of \hwplusplus\ .

\subsection{An event generator for $e^+e^-$-collisions}

\begin{figure}[h!]  
  \begin{center}
    \epsfig{file=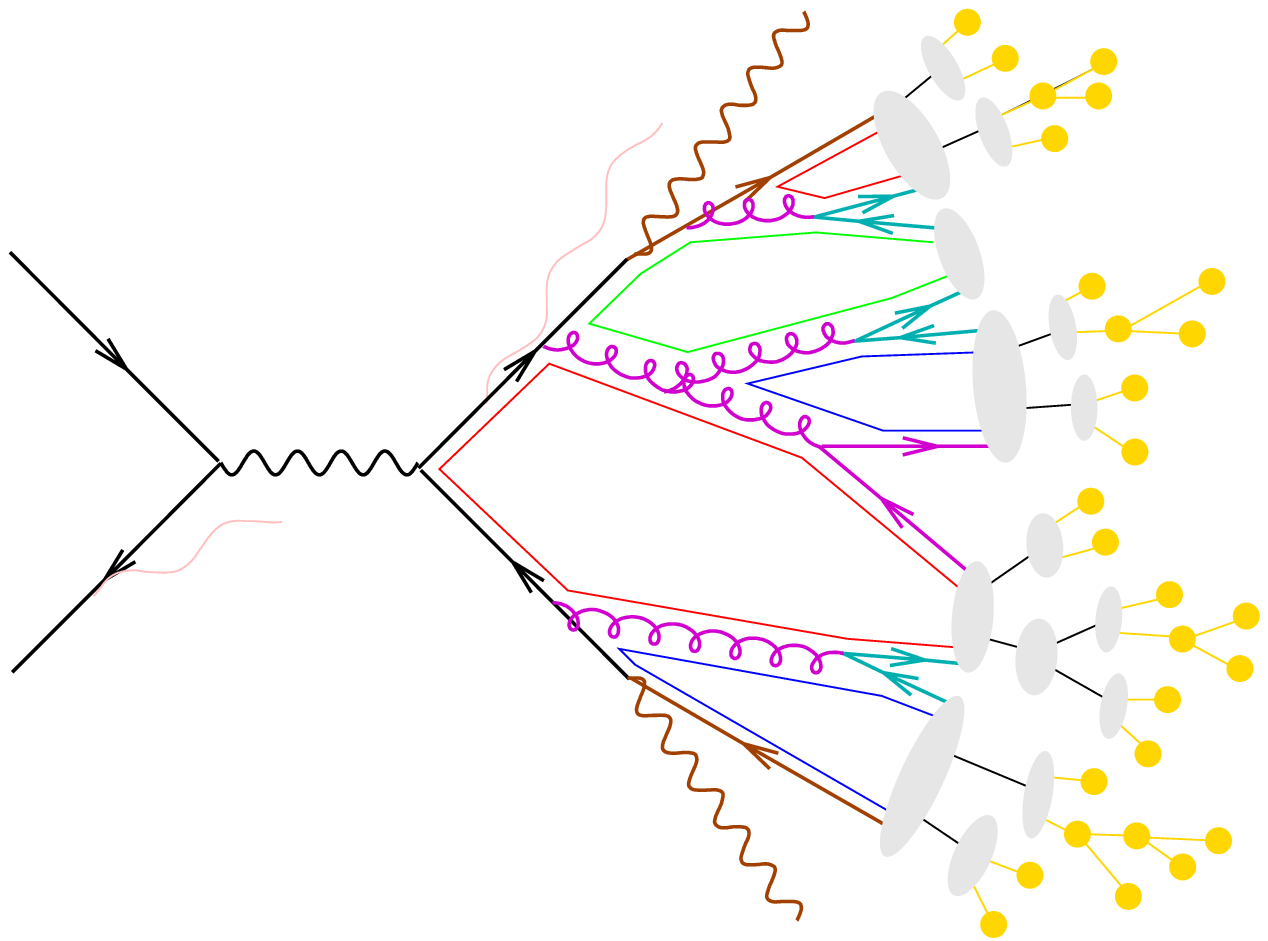,height=5.4cm,angle=90}
    \hfill
    \epsfig{file=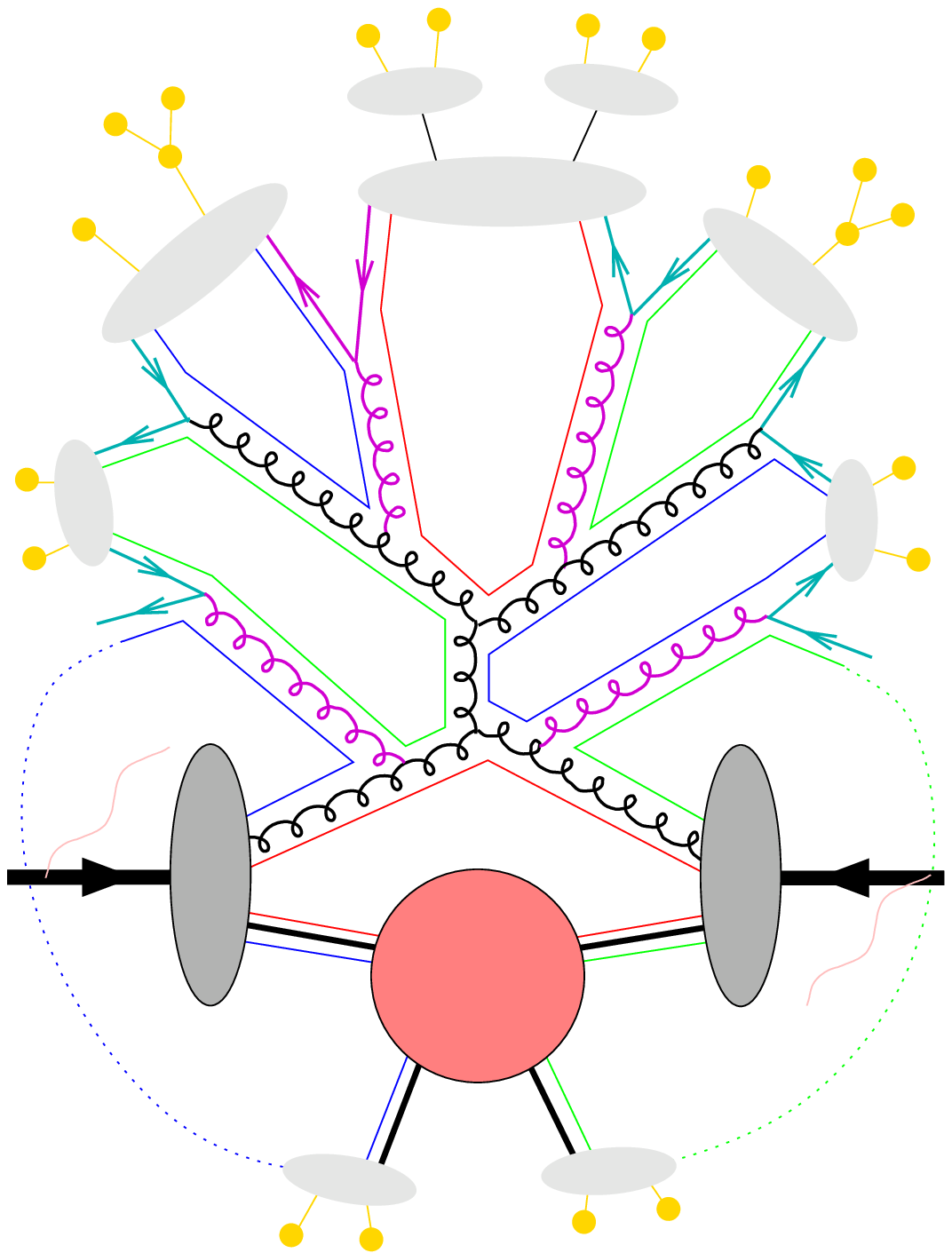,width=5.5cm}
   \end{center}
   \caption{An event generator for $e^+e^-$-collisions (left) and 
     the more complicated event structure in a $p\bar p$-collision (right)
     \label{fig:epemcoll}}
\end{figure}
The basic steps of the event generation in a Monte Carlo program for
$e^+e^-$ collisions are sketched in Fig.~\ref{fig:epemcoll} (left).
The incoming $e^+e^-$-pair generates a quark-antiquark-pair ($t\bar t$
in this example) via the exchange of a vector boson ($\gamma$ or
$Z^0$).  This central production mechanism is usually referred to as
the `hard process'.  The incoming and outgoing electrically charged
lines may additionally radiate soft photons.  Following the hard
production process the coloured particles are subject to parton shower
evolution: they radiate a number of partons, mainly gluons.
Technically this means that large logarithms from those regions in
phase space that are enhanced by soft and/or collinear emissions are
resummed.  From the point of view of the final state this means that
the coloured particles will be surrounded by a cloud of more coloured
particles that will form a jet in the final state.  The parton shower
evolution is a complete perturbative description and is terminated at
a low scale $\mu_0$ of the order of $\lqcd$.  The colour structure is
kept track of in the large-$N_c$ approximation.  This means that the
colour of gluons can be represented by a colour-anticolour-pair as if
it was a $q\bar q$-pair.

Following the perturbative description of the parton shower evolution
the partonic final state is converted into an exclusive hadronic final
state by means of a hadronization model.  Aside from two other models,
the Field-Feynman-type fragmentation model \cite{FFmodel} and the Lund
string fragmentation \cite{Lundfrag}, implemented in \pythia, the
\emph{cluster hadronization model} \cite{wolframcluhad,webbercluhad}
is a very popular one and implemented in present \herwig.  In order to
convert the partons in the final state into hadrons, the gluons in the
final state are split nonperturbatively into $q\bar q$-pairs with
colours according to the colour structure of the gluons.  Eventually,
pairs of matching colour partners can be found in the final state,
being paired up into colourless clusters that carry the sum of the
momenta of the constituent partons.  These clusters can undergo
further decays and are then converted into hadrons that carry the
appropriate quantum numbers and momenta weighted according to
available phase space and angular momentum.  These possibly unstable
hadrons decay according to well-known branching ratios such that the
majority of hadrons in the final state will be low-mass pions and
kaons.
 
\subsection{Additional complications in $p\bar p$ collisions}

In proton-antiproton collisions the above picture is significantly
complicated by the presence of coloured particles in the initial state
(cf.\ Fig.~\ref{fig:epemcoll}, right).  In addition, the initial state
particles of the hard process have to carry momentum fractions of the
initial state hadrons according to the parton distribution functions
(pdfs).  Then, these coloured initial state partons typically radiate
coloured particles as well, as described by a so-called backward
evolution which technically differs from the final-state evolution by
the presence of pdfs.

Apart from this initial state radiation the remnant particles in the
initial state might undergo further interactions, in the so-called
\emph{soft underlying event}.  Note that, as depicted in
Fig.~\ref{fig:epemcoll} (right), the colour structure is connected to
the colour structure of the final state partons.  The final state
arising from the underlying event was modeled according to a simple
model from UA5 \cite{underlyingUA5} in recent \herwig, used to describe
the $p_\perp$-distribution of minimum bias events.  Recently
\cite{Pythia,multiple}, multiple interactions above some hard scale
are taken to model the interaction of the beam remnants.

\piccaption[]{The cluster mass distribution is independent of the hard
  scale\label{fig:clumass}}
\parpic(5cm,6.6cm)[r]{
  \epsfig{file=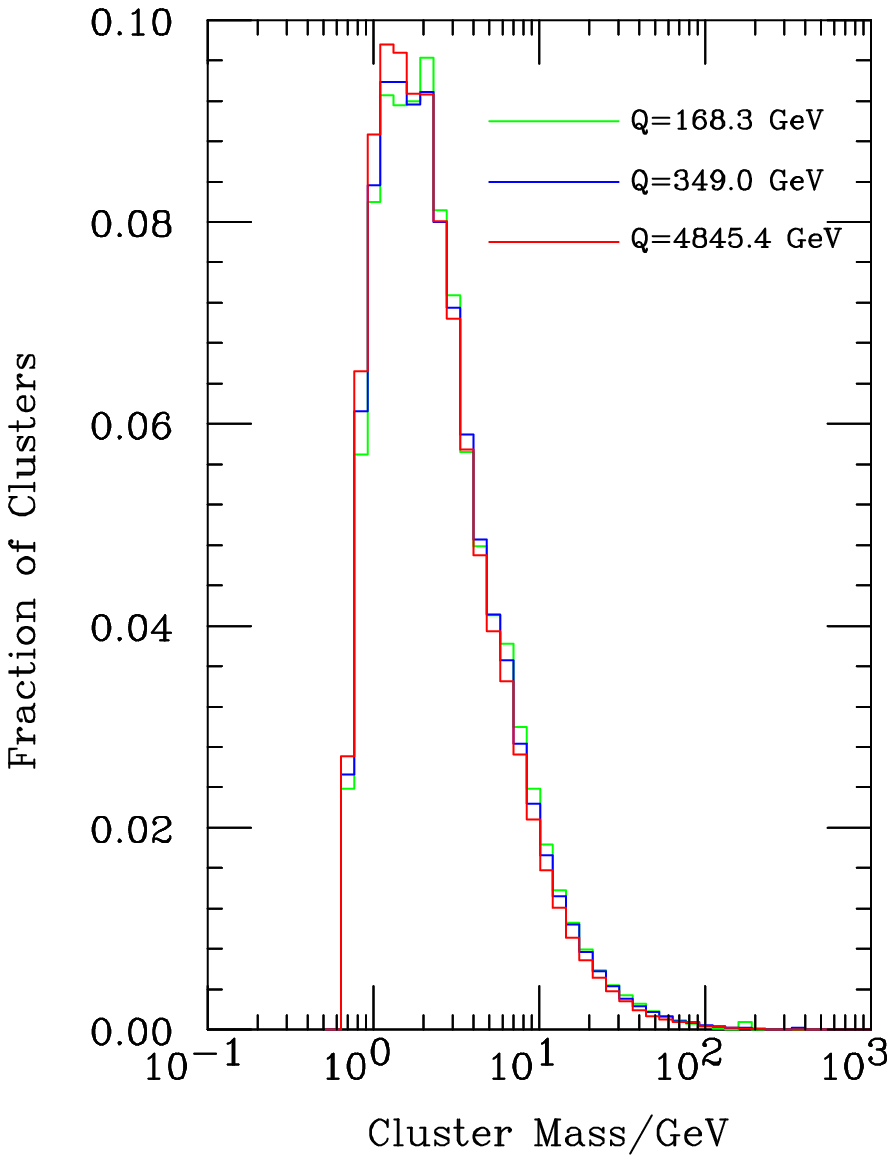,width=5cm}}
\noindent
The multi-purpose models based on such kind of model have proven to
describe the data at LEP very well in general.  However, there are
certainly observables where an increased accuracy reached by a MC
model as described above is highly desirable.  Obviously the modelling
of the ha\-dronization gives some room for improvement when observables
that depend on long-range effects are considered.  In general, the
modelling of the hadronization is not affected by the large scales in
the problem.  As shown in Fig.~\ref{fig:clumass}, the cluster mass
distribution obtained in the cluster hadronization model of \herwig\ is
independent of the centre-of-mass energy of the hard process.  This
property is well-known as colour pre-confinement in QCD. 

\section{Matrix elements and parton showers}
\label{sec:meps}

For hard scales the situation that was described above is quite
different. If one considers observables where large transverse scales
play a dominant role, the modelling of jets with the parton shower
alone fails.  Of course, the parton shower, based on the universal
soft and collinear behaviour of matrix elements in QCD is obviously
only valid in this kinematical domain of the coloured particles in the
process. 
\piccaption[]{Different orders in the LL approximation and in
    $\as$\label{fig:orders}}
\parpic(5cm,6cm)[r]{
  \epsfig{file=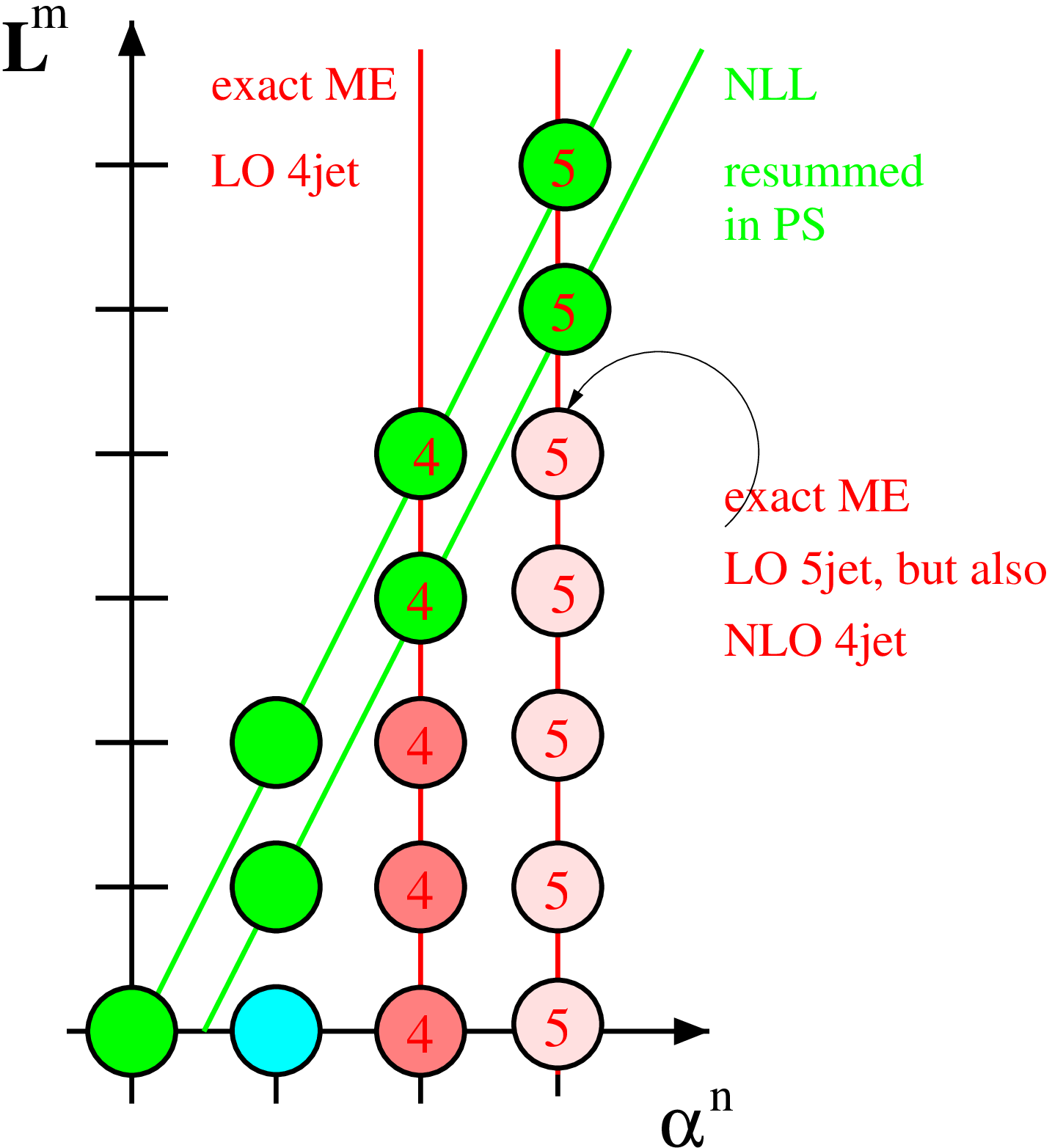,width=5cm}}  
\noindent
However, it has the advantage of describing at least the leading
contributions correctly to all orders.  In Fig.~\ref{fig:orders} this
situation is clarified for the case of multi-jet production.  Only the
top two rows on the diagonal are well-described by the parton shower
approximation.  In the following, several attemps to take the
advantages of both, fixed order approximation and the parton shower,
into account for the description of different observables are
discussed.  The aim is to keep the accuracy of fixed order
calculations as well as the possibility to produce an undetermined and
possibly large number of partons in the soft and collinear domain of
the phase space of an observable, where the parton shower
approximation is assumed to hold.

\subsection{Matrix element corrections}
\label{sec:mecorr}

\piccaption[]{The phase space for $Wg$ production\label{fig:wgprod}}
\parpic(5.1cm,4.5cm)[r]{
  \epsfig{file=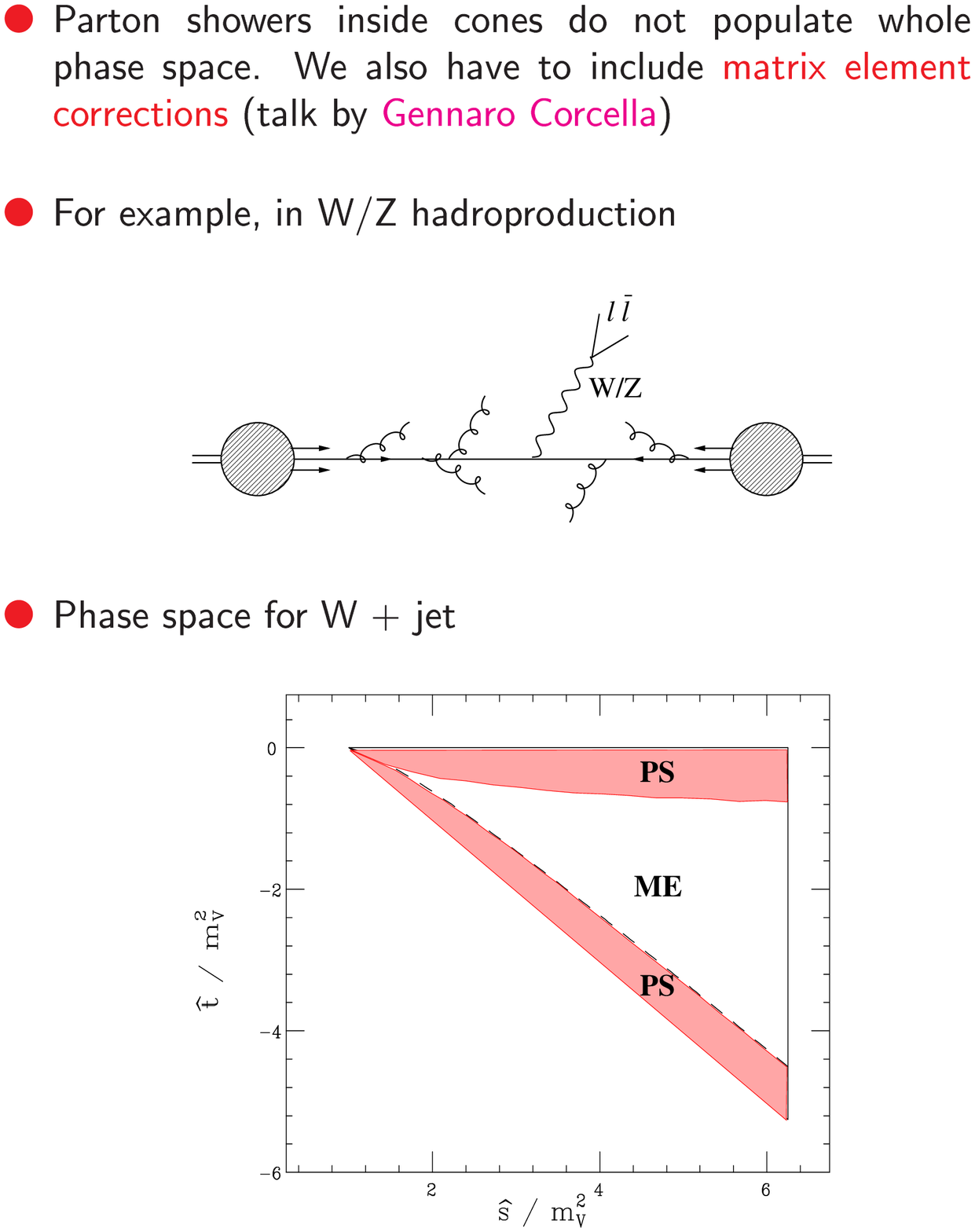,clip=true,width=5cm}} 
\noindent
In the simplest cases the above extensions are important in the decay
$t\to Wb$ or the production of vector bosons in hadronic collisions
where an extra gluon is radiated.  Usually the $g$ is generated from
parton shower radiation.  Even though the largest part of the matrix
element (ME) is described well in this approximation, this approach
fails when the gluon initiates an extra high-$p_\perp$ jet.  Here, one
can improve the description of the final state tremendously by
considering matrix element corrections
\cite{Corcella:1998rs,Seymour:1994df}.

\begin{figure}[htbp]
  \centering
  \epsfig{file=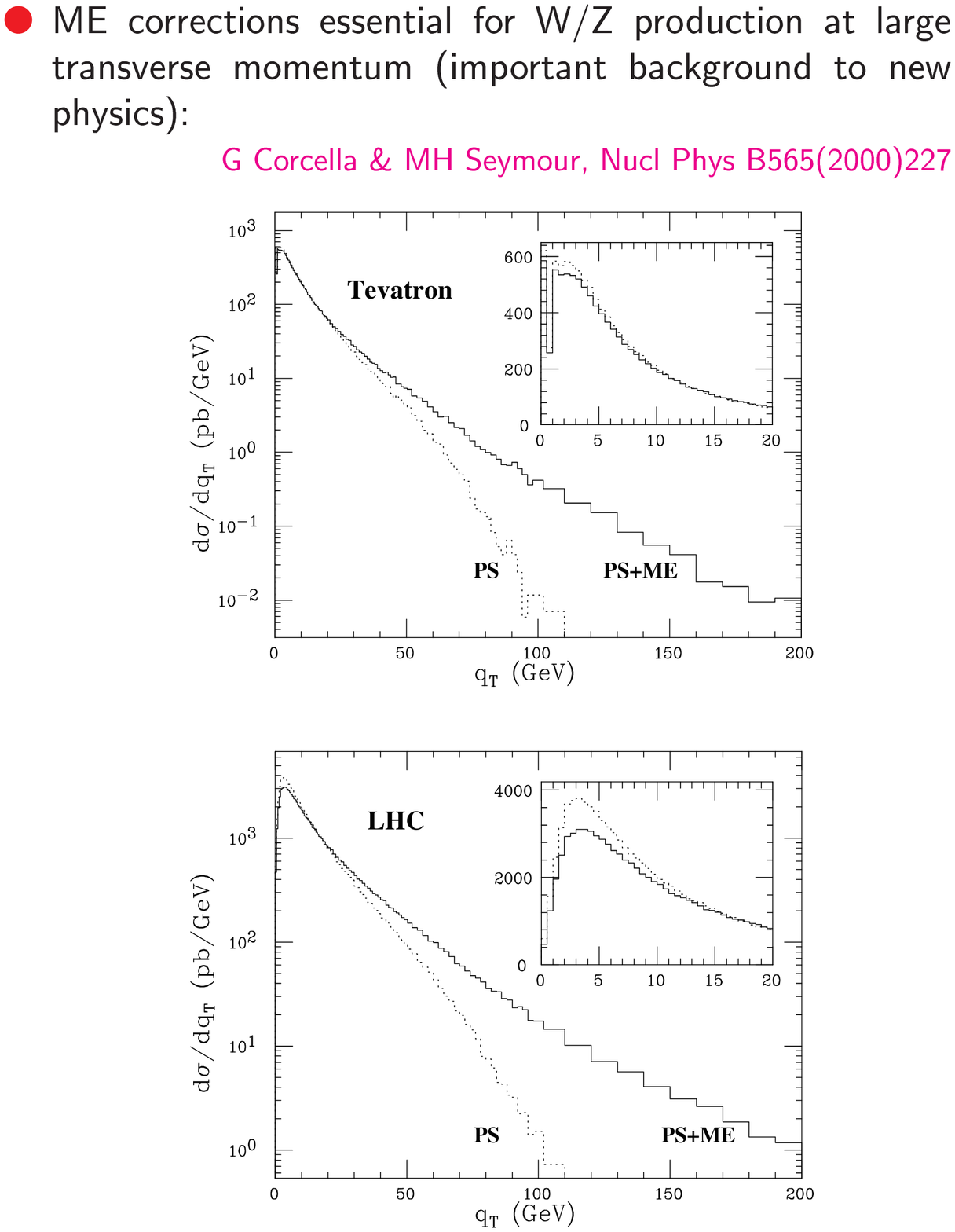,clip=true,%
    bbllx=125,bblly=400,bburx=470,bbury=665,width=5.5cm}  
  \epsfig{file=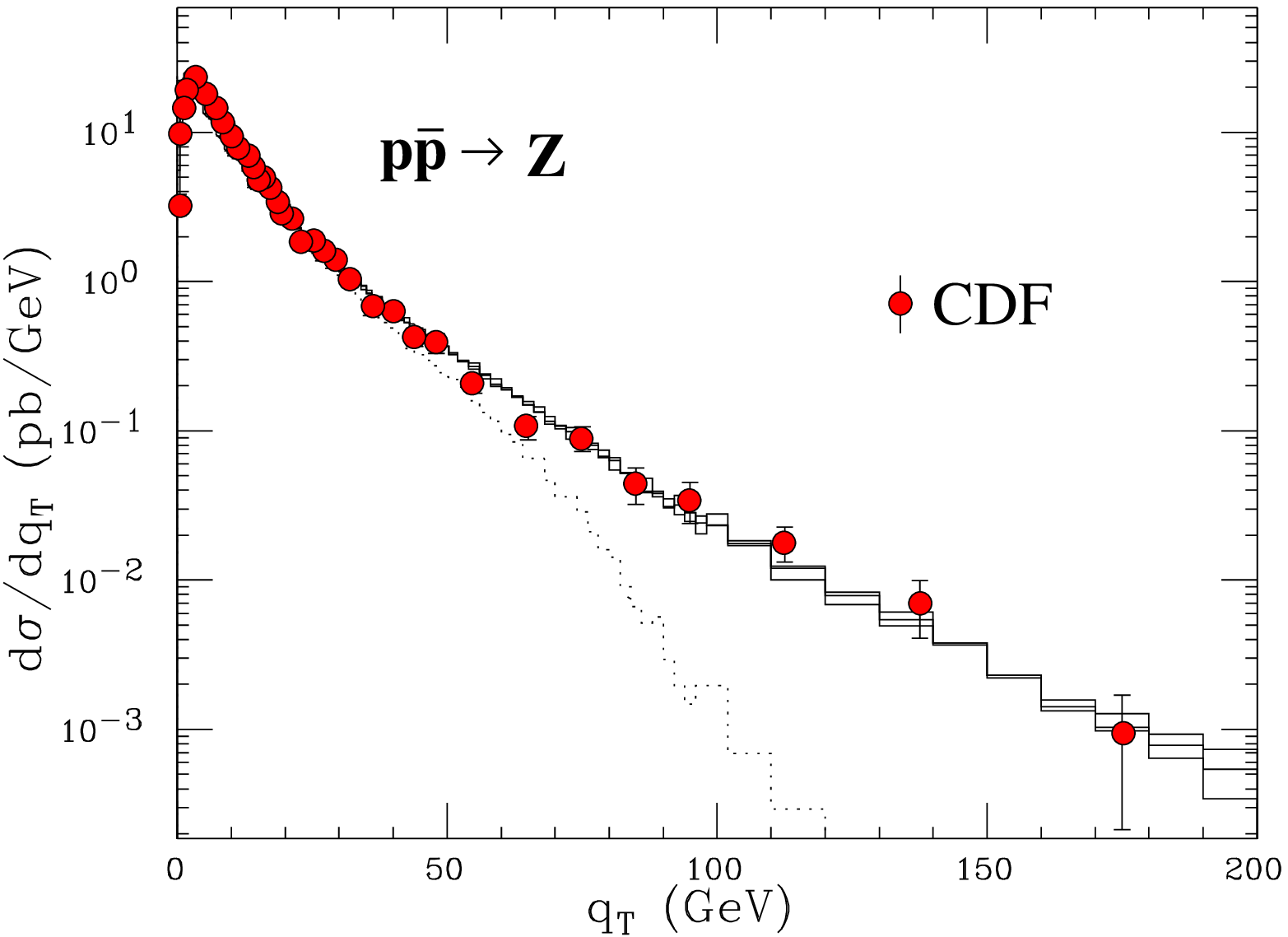,width=5.5cm}  
  \caption{$p_\perp$-distribution for $Wg$ production with and witout
    ME corrections (left).  $p_\perp$-distribution for $Zg$ production
    in comparison to CDF data (right)}
  \label{fig:mecorr}
\end{figure}

In Fig.~\ref{fig:wgprod} the phase space for $Wg$-pro\-duc\-tion in
hadron-hadron collisions is shown.  The regions `PS' are covered by
the parton shower while a hard gluon emission is typically from the
region `ME' that is not even covered by the \herwig\ parton shower in
this case.  Therefore this region is often denoted as `dead region'.
When the phase space distribution of the hardest gluon emission is
generated from the full matrix element rather than from the parton
shower, a significant part of high-$p_\perp$ radiation is added to the
parton shower result, as is shown in Fig.~\ref{fig:mecorr} for the
Tevatron.  The $p_\perp$-distribution for $Z$-production is shown in
the right panel of the same figure, together with experimental data
from CDF. 

In conclusion we note that the contributions from the matrix elements
are clearly important and the parton shower approximation might not be
sufficient for observables in which large transverse momenta play 
a r\^ole.

\subsection{Matching LO matrix elements with parton showers}
\label{sec:lomeps}

\cite{CKKW,Webber:2000mf} can be considered as the first systematic
approach that has been implemented into an event generator for
multi-jet production.  The algorithm is suitable for an undetermined
number of partons in the final state.  The number of hard jets that
can be desribed correctly by this approach is in principle only
limited by the capabilities to calculate the matrix elements for these
processes. 

The algorithm generates momenta for a given $n$-gluon final state
according to the exact matrix element.  The number of gluons has been
preselected according to a known rate with a (Durham) jet resolution
cut $y_1$.  The final state particles are then successively clustered
backwards until only the LO configuration survives.  This gives a
resolution scale at each node where two particles were clustered,
allowing to calculate a weight from the Sudakov form factors, telling
that these final states are actually unresolved down to the given
scales, and a weight taking into account the correct values of the
strong coupling.  This particular final state is then accepted
according to the calculated weight and a (vetoed) parton shower
evolution is applied to the resulting legs in between the appropriate
scales.  \cite{CKKW} have proven that the final rate is then
independent of the cut parameter $y_1$ and correct up to NLLA.  Results
for a typical 4-jet variable, the Bengtsson-Zerwas angle, are
displayed in Fig.~\ref{fig:bz} and show that this algorithm,
implemented into the event generator \apacic\ \cite{apacic} gives
results to a high precision.  We note, that this approach has been
generalized to $p\bar p$ collisions recently \cite{Krausspp}.  Another
implementation of these ideas is realized in the dipole cascade
\cite{leifmeps}.

\piccaption[]{Begtsson-Zerwas angle at LEP, compared to \pythia\ 
  and \apacic\label{fig:bz}}
\parpic(6cm,8.5cm)[r]{
  \epsfig{file=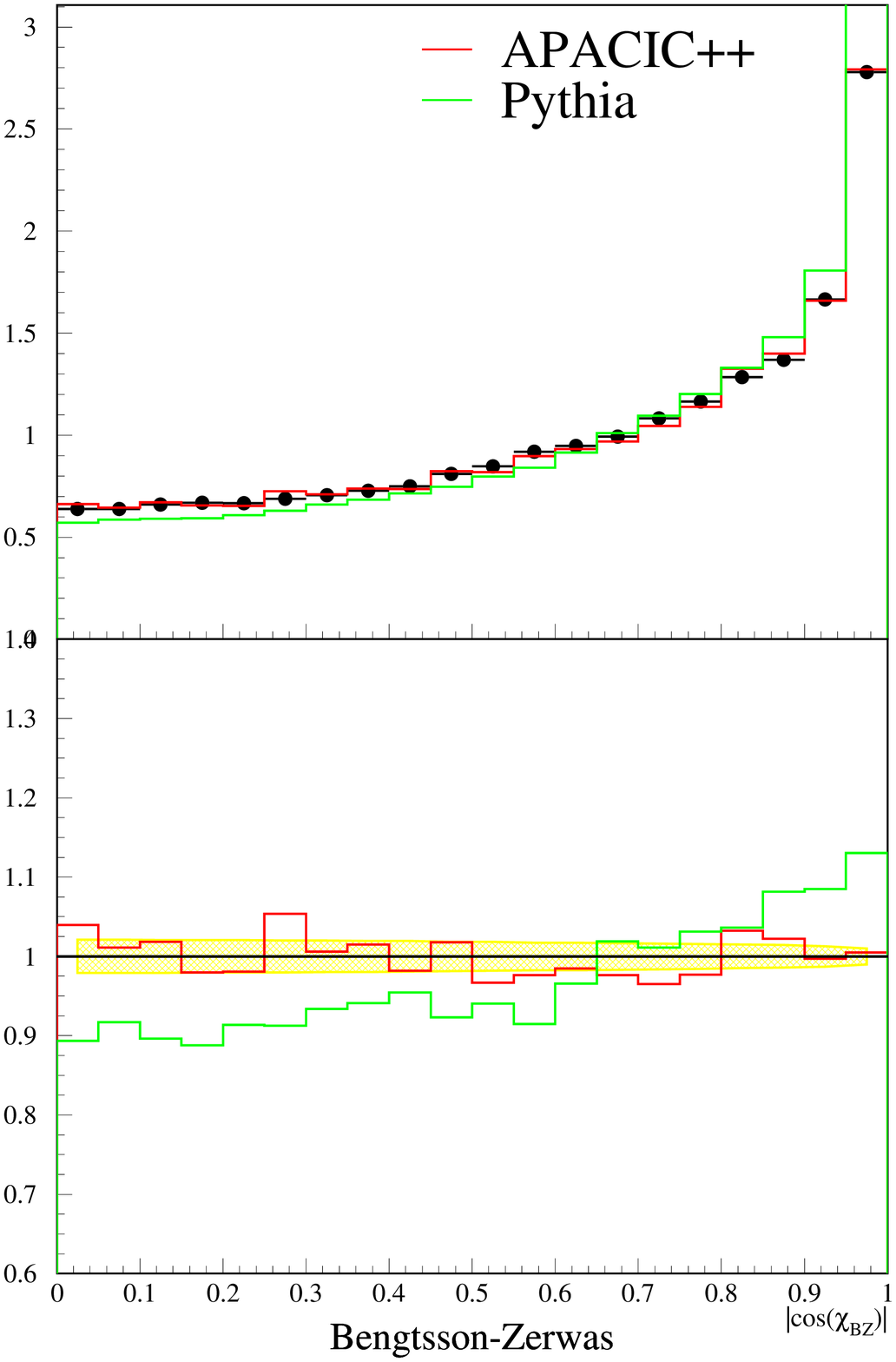,width=6cm}}
\noindent
The success of this approach raises the question of appropriate matrix
element generators that allow for a generation of multijet final
states to a very high precision, to be combined with the benefits of
the Monte Carlo event generator to generate additional LL
QCD-radiation and to produce an exclusive final state.  Two recent
examples are \amegic\ \cite{amegic} and \madevent\ \cite{madevent}.
The latter allows for a rather robust generation of final states with
a large number of jets while the former aims for high precision at the
same time.  Both are multi-purpose programs that do not require any
user interaction apart from entering the desired process, of course.
Even this task can in principle be done automatically.  Other examples
of multi-purpose matrix element generators are \alpgen\ \cite{alpgen}
and \comphep \cite{comphep}.

\subsection{Matching parton showers with NLO matrix elements}
\label{sec:mcnlo}

\piccaption[]{Results from \cite{FW} for $W$-pair production
  \label{fig:fwww}}
\parpic(6.2cm,4.5cm)[r]{
  \epsfig{file=wwptpair.eps,width=6cm}}
\noindent
In addition to aiming for a large number of jets in the final state,
improved precision is achieved by taking into account NLO matrix
elements as well.  The challenge is to correctly match up the emission
of an extra gluon from the real correction with the the parton shower
emission.  Three groups have investigated this problem in different
ways.  \cite{dobbs} and \cite{poscho} have used the phase space
slicing method to separate hard and soft (unresolvable) emissions with
good success in the description of e.g.\ the differential jet shape at
HERA.  \cite{FW}, however, argue that their results suffer from an
inconsistency: they do not reproduce the perturbative orders at any
point in phase space correctly.  In contrast, the latter use a
subtraction method that suits the needs of the Monte Carlo program
which does not have to be modified to a large extent.  The
implementation \mcnlo\ \cite{mcnlo} generates results for $W$-pair
production that show the desired features (Fig.~\ref{fig:fwww}): at
high $q_\perp$ the program matches up the NLO result while at low
$q_\perp$ the soft and collinear divergences are properly resummed by
the Monte Carlo program.
The group \cite{collins} has achieved important theoretical results as
well as matching of NLO matrix elements to parton showers for certain
processes.

\section{Development of \hwplusplus}
\label{sec:hw++}
In this final section the status of the development of the new Monte
Carlo event generator \hwplusplus\ is outlined.  As the name might
suggest, \hwplusplus\ is a new C++ version of the well-known event
generator \herwig.  However, \hwplusplus\ will not be a plain
`rewrite' of the same program but will have some new features added.
There are several major motivations for a complete rewrite in another
programming language.

First of all, the existing Fortran code has been constantly developed
throughout nearly twenty years and is maintained by a large number of
authors making it increasingly difficult to maintain the program
efficiently and to guarantee the safety of code, i.e.\ it is in
principle possible that one author modifies parts of the program which
he did not intend to touch at all.  In addition, more and more users
wish to modify the code themselves, adding for example matrix elements
they obtained from new models.  Aside from others, these examples
clearly show the requirements that a new version of the program should
fulfil.  The object oriented features of C++ perfectly match these
requirements.

Some major arguments in favour of such a rewrite are: 
\begin{itemize}
\item The experimental collaborations start to write new (and to
  rewrite existing) analysis software entirely in C++.  Since the
  program will be used by people working in these collaborations it is
  much more likely that a succesful process of bug report and handling
  is achieved when the users actually know the language a program is
  written in very well.
\item Benefit along similar lines is achieved since C++ has become the
  standard programming language in the UNIX/LINUX world, which is
  clearly dominating in the HEP world.
\item Object oriented code is much easier to maintain since parts of
  the code are \emph{encapsulated} and therefore the modification of
  one part of the program cannot affect another part of the program.
\item The maintenance is easier for similar reasons.  Even after a
  long period one might modify the code for a specific process without
  having to know all the details of the remainder of the program.
\item For similar reasons it will be easy to implement code for new
  physics processes.  The user does not have to be worried about
  modifying code he did not intend to.
\end{itemize}

\begin{figure}[htbp]
  \begin{center}
    \epsfig{file=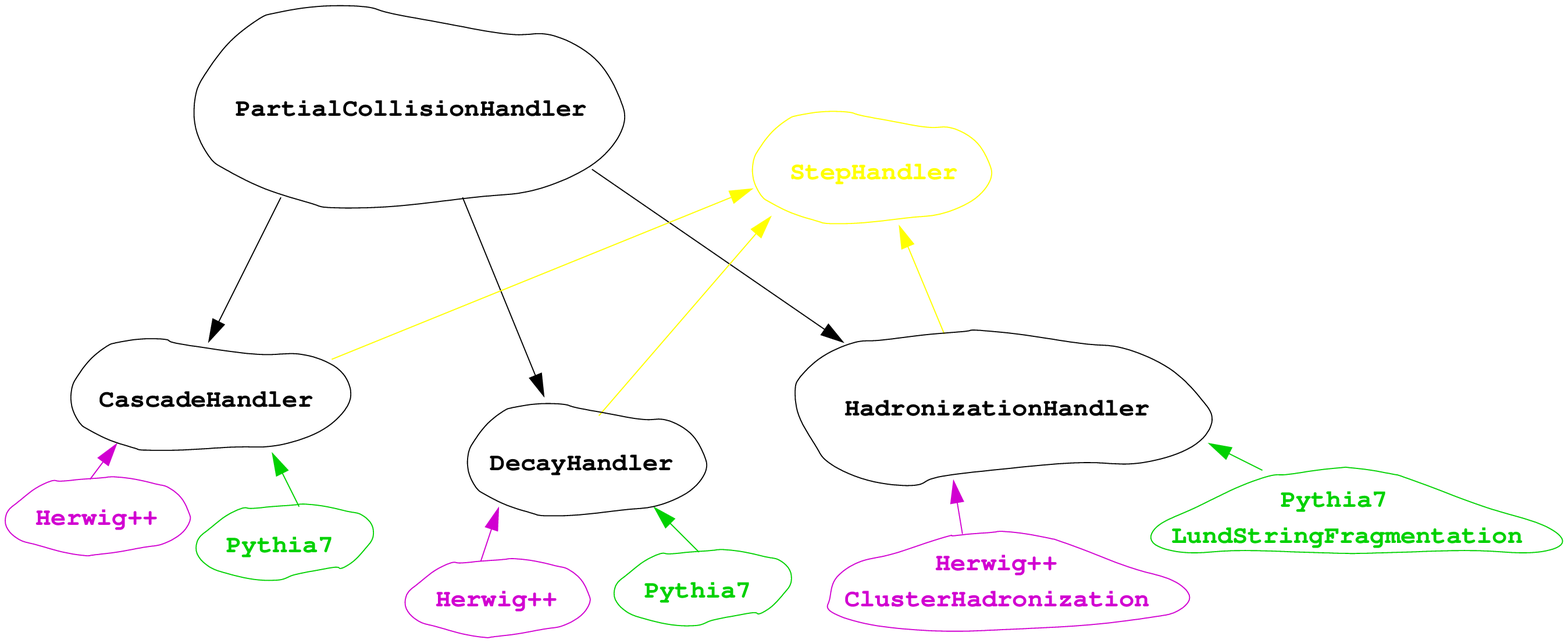,width=\textwidth}
    \caption{The PartialCollisionHandlers in \hwplusplus\ and 
      their interplay with \p7}
    \label{fig:hwp7}
  \end{center}
\end{figure}
Apart from a different programming language, \hwplusplus\ will also
depend on parts of the new program \p7 \cite{p7}, a similar project
for the well-known program \pythia.  One must note that \p7\ consists
of two major pieces.  First of all a library of very general classes
was developed that contains many useful features.  On top of this
library there will be the physics implementation of the event
generator \pythia.  Upon the start of the \hwplusplus\ project the
library part of \p7\ was already completed and it was decided to build
the physics implementation of \hwplusplus\ on top of this library.
The way in which \hwplusplus\ hooks into the structure of \p7\ is
sketched in Fig.~\ref{fig:hwp7}. Even though these classes were
designed very properly and the whole approach was intended to be as
general as possible, one might expect the disadvantage that the two
models are not completely independent anymore.  However, on the other
hand, users of both models might benefit from the common environment
in several ways.  Among these benefits there are the possibility to
use parts of both programs, e.g.\ \hwplusplus's parton shower together
with the Lund string fragmentation from \p7.  Furthermore, one might
have a common graphical user interface.

In summary, the idea is to have a large and flexible, object-oriented
implementation of the physics models that are implemented so far in
the Fortran version \herwig\ with the possibility of a straightforward
extension and adaptation to future requirements.  In the following,
two major physics improvements in the parton shower will be described
in greater detail.

\subsection{New parton shower variables}
\label{sec:newvar}

\begin{figure}[htbp]
  \begin{center}
    \epsfig{file=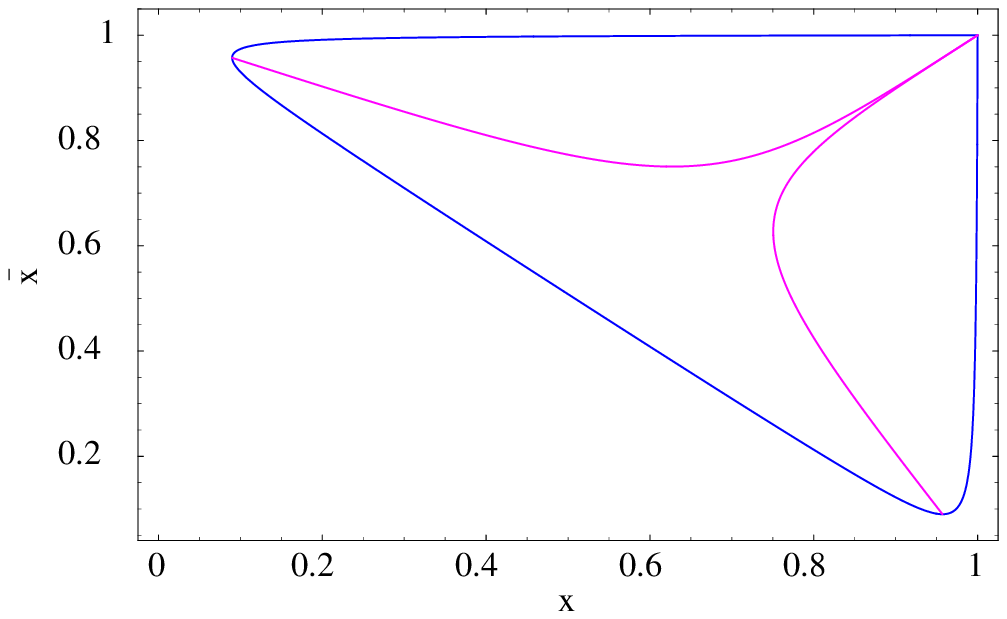,width=5.5cm}
    \epsfig{file=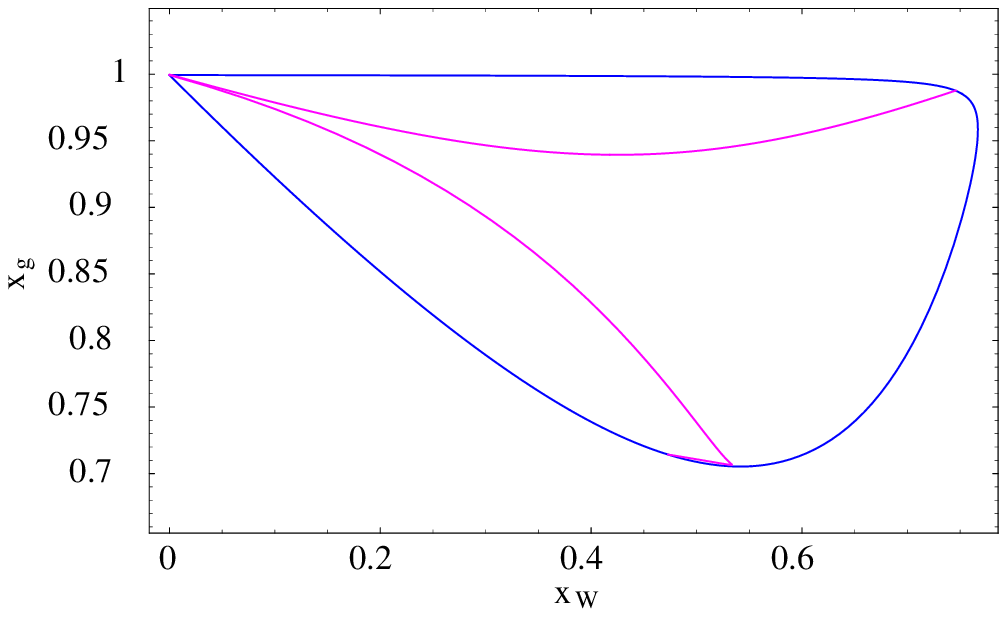,width=5.5cm}    
    \caption{Phase space covered by the parton shower in $e^+e^- \to q\bar qg$ (left) and $t\to Wbg$ (right)}
    \label{fig:psnew}
  \end{center}
\end{figure}

\noindent
One major improvement of the parton shower in \hwplusplus\ will be the
usage of a new set of evolution variables \cite{cacc_cata,newvar}.
The previous \herwig\ evolved partons in an angular variable $\xi$,
suffering from complicated `dead' cones and an overlap in the soft
region of the 'final state+gluon' phase space.  It is possible to
overcome these problems with a new evolution variable $\tilde q^2$.
Based on a Sudakov basis $p, n$ with $p^2 = m^2$, $n^2=0$ we decompose
the partons' momenta in the shower as shown in
Fig.~\ref{fig:pskin}\piccaption[]{Kinematics of the parton
  shower\label{fig:pskin}} \parpic(5cm,3.5cm)[r]{
  \epsfig{file=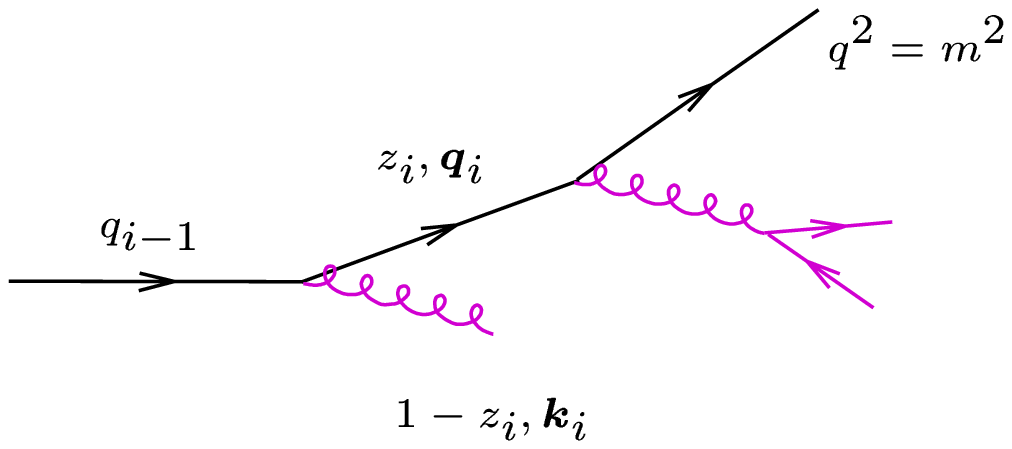,bbllx=104,bblly=569,bburx=391,bbury=695,clip=yes,width=5cm}}
\begin{displaymath}
  q_i = \alpha_i p + \beta_i n + \bm q_i. 
\end{displaymath}
The vector $p$ is the momentum of the jet's parent particle and $n$
defines a backward direction in a suitable way.  The longitudinal
splitting is then defined relative to the $p$-direction,
\begin{displaymath}
  \alpha_i = z_i \alpha_{i-1}. 
\end{displaymath}
The evolution variable 
\begin{equation} 
  \tilde q^2 = \frac{\bm p^2}{z^2(1-z)^2} + \frac{m^2}{z^2},
\end{equation}
with the argument of running $\as$ chosen according to 
$\as(z^2(1-z)^2\tilde q^2)$, determines the transverse momenta 
via $\bm p_i$, 
\begin{equation}
  \bm q_i = \bm p_i + z_i \bm q_{i-1}, 
  \qquad \bm k_i = - \bm p_i + (1-z_i) \bm q_{i-1}\,.
\end{equation}
Having chosen an azimuthal angle $\varphi$ still randomly or as a
result of planned azimuthal spin correlations \cite{peterspin} we can
reconstruct the kinemaics recursively from the onshellness of the
final state particles in the shower. Note that angular ordering is
satisfied in terms of
\begin{equation}
  \tilde q_{i+1} < z_i \tilde q_i,  \qquad \tilde k_{i+1} < (1-z_i)
  \tilde q_i\,.
\end{equation}
Technically, the new evolution variables only lead to a
reinterpretation of the well-known Sudakov form factors, i.e.\ the
branching probability for parton $a$ splitting to partons $bc$ is
still given in terms of the usual splitting function as
\begin{equation}
  dP(a\to bc) = \frac{d\tilde q^2}{\tilde q^2}
  \frac{C_i\as}{2\pi} 
  P_{ba}(z)\,dz
\end{equation}
where $C_i$ is a colour factor.  Considering the two processes $e^+e^-
\to q\bar qg$ and $t\to Wbg$ in Fig.~\ref{fig:psnew} it is clear that
a smooth coverage of the soft region can be achieved by choosing
appropriate limiting values for the parton shower.  This allows for
the generation of soft radiation without double counting.

\subsection{Multiscale shower}
\label{sec:multi}

The next major improvement of the parton shower will be the
implementation of a \emph{multiscale parton shower} algorithm that
takes the evolution of unstable particles and their widths properly
into account.  In order to outline the basic features of such a
showering algorithm let us briefly recapitulate the treatment of the
parton shower evolution of the $t$-quark in recent \herwig.
\begin{figure}[htbp]
  \begin{center}
    \epsfig{file=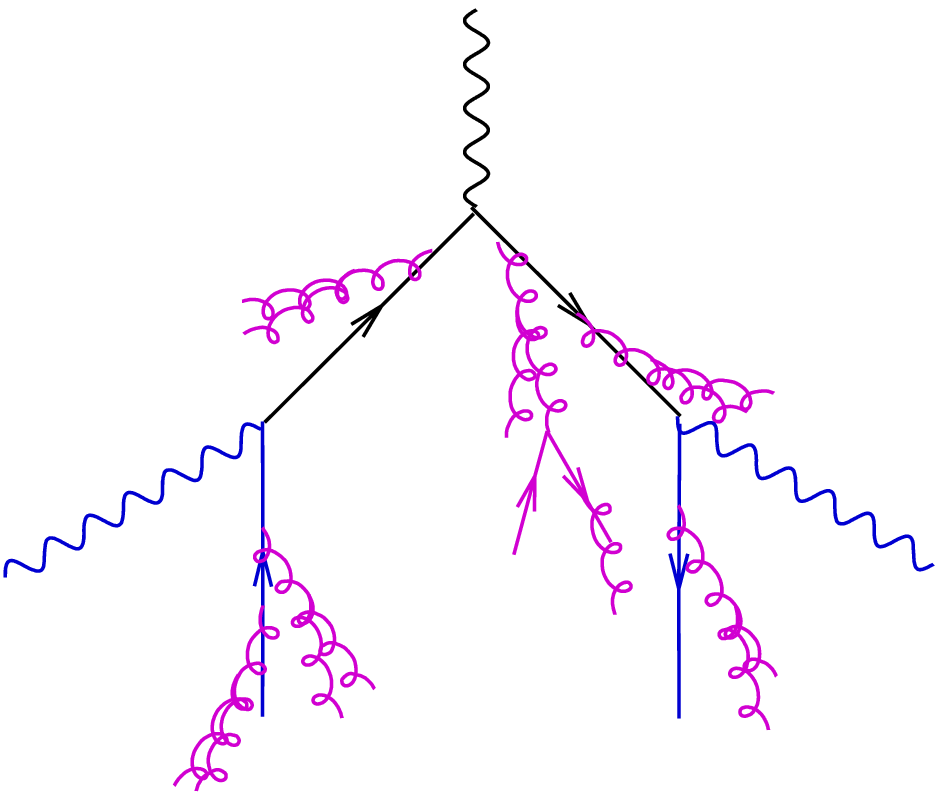,height=5.2cm,angle=270}  
    \hfill
    \epsfig{file=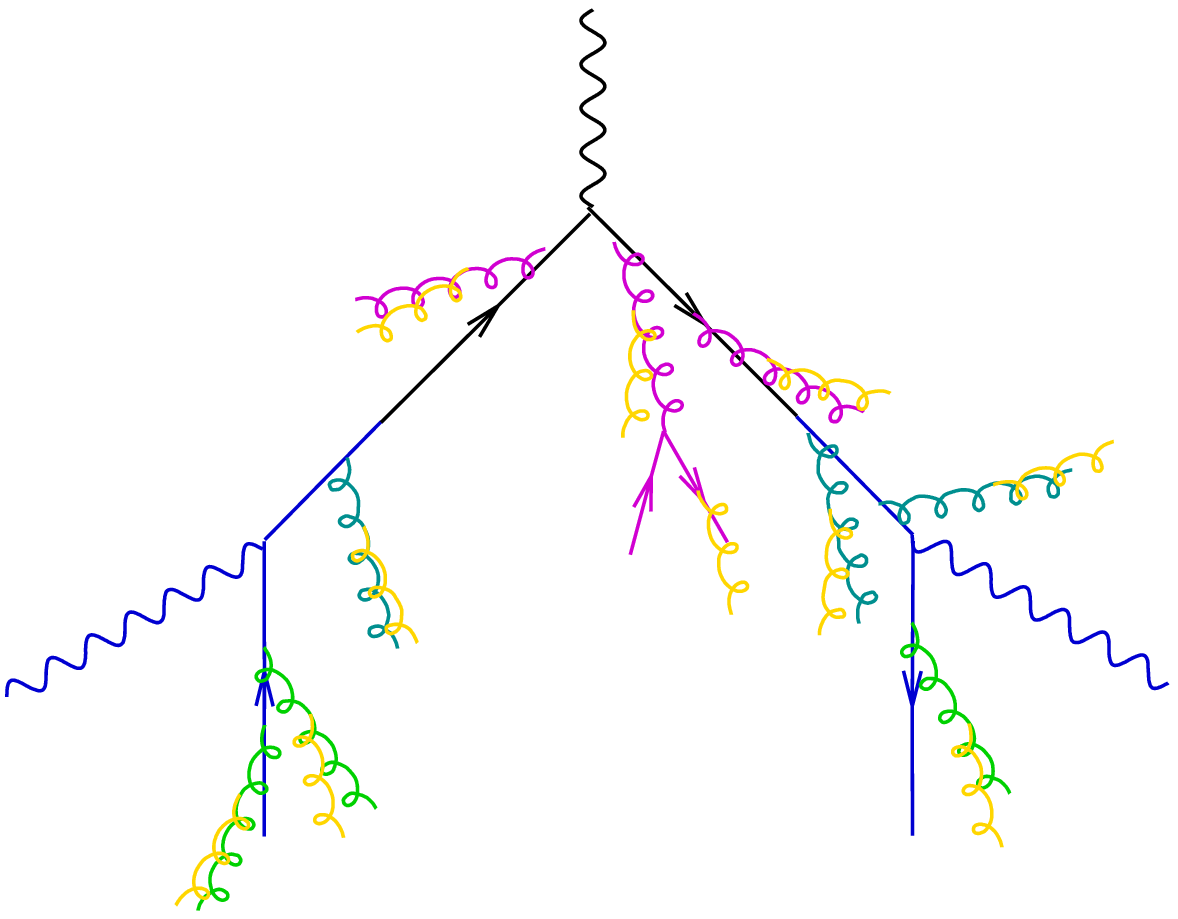,height=5.2cm,angle=270}
    \caption{Treat of QCD radiation in recent \herwig\ (left) 
      and the new multiscale shower algorithm (right)}
    \label{fig:multiscale}
  \end{center}
\end{figure}
In Fig.~\ref{fig:multiscale} (left) we have a $t\bar t$-pair produced
at a scale $\hat s$.  First there will be the evolution of the
top-quarks between the scales $(\hat s \to \mu_0)$ with subsequent
decays of the $t$-quarks, $(t\to Wb, \bar t\to W\bar b)$ and finally
the parton shower evolution of the $b$ ($\bar b$) within $(m_t \to
\mu_0)$.  In both cases we have QCD-radiation down to a lowest
resolution scale $\mu_0$, ignoring the finite width $\Gamma_t$ of the
$t$-quark.

In the Multiscale Shower, cf.\ Fig.~\ref{fig:multiscale} (right), on
the other hand, the $t$-quarks will only be evolved down to their
width ($\hat s\to \Gamma_t$) before they decay. Next, they decay
($t\to Wb$, $\bar t\to W\bar b$) and a (DIS-like) backward-evolution
($m_t\to\Gamma_t$) from $t$, $\bar t$ is generated to take into
account that the $t$ quarks are actually off-shell, quite in contrast
to the description of the hard matrix elements, where they are assumed
to be on-shell. Next, the $b$, $\bar b$ are evolved $(m_t\to\Gamma_t)$
down to the lowest scale taken into account so far. Finally, the
evolution from $(\Gamma_t\to\mu_0)$ is generated \emph{globally}.  In
case also the $b$ were unstable, the last global evolution would have
been going down to only $\Gamma_b$ and the algorithm would have
carried on as in the $t$-case.

In conclusion, the multiscale shower algorithm will be suitable for a
description of the QCD evolution in complicated decay chains, as they
might become important e.g.\ for supersymmetric particles.  The
algorithm is general enough to deal with arbitrary width and masses,
assuming the usual collinear factorization. The classes for the
multiscale shower are already implemented in \hwplusplus\ and the
shower algorithms are designed appropriately even though the
`concrete' code is not yet completed.  Old \herwig\ results should be
reproduces with a simple set of switches. Furthermore, we note that
also different types of radiation are foreseen in the shower design
even though their physical relevance remains to be clarified.

\subsection{Status of the program}
\label{sec:stat}

As to the remaining parts of the program, we briefly summarize their
status of development.  The partonic decays are clearly interwoven
with the parton shower as it should be clear from the previous
section.  In order to be able to treat the emission of multiple gluons
with high precision, if needed, we have implemented and tested an
interface to the matrix element generator \amegic\ for the case of the
top quark decay.  The vast amount of hadronic decays will be treated
as in \herwig\ in a first version.  This may be very useful for
testing purposes since as a first step we want to reproduce results
from \herwig\ with \hwplusplus.  In addition, interfaces to existing
decay tables e.g.\ from \EvtGen\ and \geant4\ are foreseen.  Hard
processes will be treated in a similar way: the important $2\to 2$
matrix elements will be hard-wired into the code while we will have
the option to have interfaces to matrix element generators like
\amegic\ in order to treat multijet production with high precision if
desired.  Again, the object oriented approach will allow users a safe
implementation of their own matrix elements.  In fact, this will be
encouraged and one might possibly think of a kind of library here.
Finally, the cluster hadronization model of recent \herwig\ has been
implemented and tested successfully along the lines of the existing
Fortran code.

In conclusion, a running version for $e^+e^-$ collisions with similar
features as in recent \herwig\ will appear quite soon.  Further stages
of development include the complete implementation and testing of the
multiscale shower algorithm and the quickest possible extension
towards the description of hadronic collisions.  The final aim is, of
course, to provide a full featured version in time for the advent of
LHC.

\section*{Acknowledgments}
It is a pleasure to thank the organisers for their invitation to this
enjoyable conference.  I would like to thank Frank Krauss, Peter
Richardson and Bryan Webber for their valuable comments and
contributions to this talk.  I am grateful to my colleagues in the
\hwplusplus-team, Alberto Ribon, Mike Seymour, Phil Stephens and Bryan Webber
for a very pleasent and fruitful collaboration.


\end{document}